\documentclass[a4paper,fleqn,usenatbib]{mnras}

\usepackage{newtxtext,newtxmath}
\usepackage{color}
\usepackage{graphicx}	% Including figure files
\usepackage{amsmath}	% Advanced maths commands
\usepackage{amssymb}	% Extra maths symbols

\usepackage[T1]{fontenc}
\usepackage{ae,aecompl}

\title[LAP photoemission and EUV]{Rosetta photoelectron emission and solar ultraviolet flux at comet 67P}

\author[F. L. Johansson et al.]{Fredrik L. Johansson,$^{1,2}$\thanks{E-mail: frejon@irfu.se (FLJ)}
E. Odelstad,$^{2,1}$
J. J. P. Paulsson,$^{3}$
S. S. Harang,$^{3}$
\newauthor %  
A. I. Eriksson,$^{1}$
T. Mannel,$^{4,5}$
E. Vigren,$^{1}$
N. J. T. Edberg,$^{1}$
W. J. Miloch,$^{3}$
\newauthor % 
C. Simon Wedlund,$^{3}$
E. Thiemann,$^{6}$
F. Eparvier,$^{6}$
L. Andersson$^{6}$
\\
$^{1}$Swedish Institute of Space Physics, Box 537, SE-75121 Uppsala, Sweden\\
$^{2}$Department of Physics and Astronomy, Uppsala University, Box 516, SE-75120, Sweden\\
$^{3}$Department of Physics, University of Oslo, Sem S\ae lands vei 24, Postbox 1048,
0317 Oslo, Norway\\
$^{4}$Space Research Institute, Austrian Academy of Sciences, Schmiedlstrasse 6, 8042 Graz, Austria\\
$^{5}$Physics Institute, University of Graz, Universit\"atsplatz 5, 8010 Graz, Austria\\
$^{6}$Laboratory for Atmospheric and Space Physics, University of Colorado, 3665 Discovery Drive Boulder, USA
}

\date{Accepted XXX. Received YYY; in original form ZZZ}

\pubyear{2017}

\begin{document}
\label{firstpage}
\pagerange{\pageref{firstpage}--\pageref{lastpage}}
\maketitle

% Abstract of the paper
\begin{abstract}
The Langmuir Probe instrument on Rosetta monitored the photoelectron emission current of the probes during the Rosetta mission at comet 67P/Churyumov-Gerasimenko, in essence acting as a photodiode monitoring the solar ultraviolet radiation at wavelengths below 250 nm. We have used three methods of extracting the photoelectron saturation current from the Langmuir probe measurements. The resulting dataset can be used as an index of the solar far and extreme ultraviolet at the Rosetta spacecraft position, including flares, in wavelengths that are important for photoionisation of the cometary neutral gas. Comparing the photoemission current to data measurements by MAVEN/EUVM and TIMED/SEE, we find good correlation when 67P was at large heliocentric distances early and late in the mission, but up to 50 percent decrease of the expected photoelectron current at perihelion. We discuss possible reasons for the photoemission decrease, including scattering and absorption by nanograins created by disintegration of cometary dust far away from the nucleus.

%{\color{red} Is this true? Brief flares would not go in -- they will be outliers in the dI/dV and direct sweep method, and hit the shadow method only at extremely good luck. In practice we will miss many flares, at least their peaks}{\color{green} there are many broad peaks in the EUV flare data, see \citep{thiemann_time_2017}. Minutes, hours //F}, 

% * <frejon@irfu.se> 2017-03-31T14:14:29.829Z:
% 
% > A large distance from the comet.
% Anders, Even if we get a result for the size and particle distribution, LAP cannot distinguish between  nanograins from afar or from very close. Only our assumptions make that distinction, so I think it's best to leave that out.  
% 
% ^.
% * <cyril.simon.wedlund@gmail.com> 2017-03-30T23:10:25.982Z:
% 
% > The resulting dataset can be used as an index of the solar EUV at the Rosetta spacecraft position, including flare
% 
% Regarding flares, I find that the discussion in the discussion section is lacking some meatball. I think a specific paragraph about flare detection, how they manifest in the data, at which temporal scales (of emission and of photoionisation), and which flares better suited (in terms of temporal resolution, C, M, or X and sensitivity) is needed. I would also suggest that you quote a few papers on the average length of flares. I remember that my former lab in Belgium had some page dedicated to this: http://www.stce.be/news/332/welcome.html
% I also remember that they were some peer-reviewed study on that too, cannot find it just now but a simple search should suffice. 
% 
% ^.
\end{abstract}

% Select between one and six entries from the list of approved keywords.
% Don't make up new ones.
%Comets -- Langmuir Probe -- photoemission -- Rosetta -- EUV -- dust

%%%%%%%%%%%%%%%%%%%%%%%%%%%%%%%%%%%%%%%%%%%%%%%%%%

%%%%%%%%%%%%%%%%% BODY OF PAPER %%%%%%%%%%%%%%%%%%

\section{Introduction}

ESA's comet-chaser Rosetta arrived at comet 67P/Churyumov-Gerasimenko in August 2014 and completed its mission in September 2016. During all this time, the instruments of the Rosetta Plasma Consortium (RPC) were monitoring the plasma environment. The Langmuir probe instrument (RPC-LAP), described in detail by \citet{eriksson_rpc-lap:_2007}, measures the current between the probe and surrounding space with the aim to characterise the plasma. When the probes are sunlit, they also measure the current due to excitation and emission of electrons from light, as discovered by \citet{hertz_ueber_1887} and famously interpreted by \citet{einstein_uber_1905}.
%The Langmuir probes of the Rosetta Plasma Consortium (RPC) measure current to and from the probe to investigate the plasma surroundings but also the photoemission current from the probe as sunlight excite and emit electrons on the titanium Nitride (TiN) coated titanium probes. 
The photoemission saturation current of a Langmuir probe depends on the solar far and extreme ultraviolet spectrum, and has successfully been used as a proxy for the solar UV flux on previous studies around Venus \citep{brace_solar_1988,hoegy_how_1993}. For plasma science, the UV flux has implications on spacecraft  charging, as well as a fundamental source for plasma from the ionisation of neutrals around e.g. a comet \citep{vigren_modelobservation_2016,vigren_predictions_2013,bodewits_changes_2016,galand_ionospheric_2016}. The photoemission, like the solar flux, should follow a $r^{-2}$ relation as Rosetta approaches and retreats from comet perihelion.

% and around Earth \citep{eriksson_photoemission_2007}

We compare the photoemission observed by RPC-LAP from May 2014 to the end of mission in September 2016 to the expected photoemission using UV observations from the SEE experiment \citep{woods_solar_2005} on the TIMED spacecraft orbiting Earth and the EUVM experiment \citep{eparvier_solar_2015} on the Maven mission at Mars. The orbits of Mars and 67P were such that MAVEN and Rosetta were on the same side of the Sun during most of the Rosetta mission, although there are times when Rosetta is better aligned with the TIMED/SEE at Earth, with the added benefit of superior wavelength resolution.

%We will compare our results with estimates of the expected photosaturation current of a sunlit probe using both the MAVEN/EUVM and TIMED/SEE data, propagated to the Rosetta position.

%This is not the case for Earth and 67P, although there are times where Rosetta alignment was better with TIMED/SEE at Earth than with MAVEN at Mars. TIMED/SEE also has better wavelength resolution than MAVEN/EUVM, so we use both datasets. 

We use three independent techniques for estimating the maximum photoemission current of a negatively charged probe, the photosaturation 
current ($I_{ph0}$), one of these techniques is to our knowledge new.

%Using observations of the UV spectrum from the SEE experiment \citep{woods_solar_2005} on the TIMED spacecraft orbiting Earth and the EUVM experiment \citep{eparvier_solar_2015} on the MAVEN mission at Mars, we compare the photoelectron emission observed by RPC-LAP from May 2014 to the end of the mission in September 2016. We use three independent techniques for estimating the maximum photoemission of a negatively charged probe, the photosaturation ($I_{ph0}$) current, one of these techniques is to our knowledge new.

In Section~\ref{sec:methods}, we go through the relevant theory of Langmuir Probes and the photoelectric effect, followed by a description of each technique to obtain the photoemission current from the probes as well as the estimates from UV observations. In Section~\ref{sec:obs} we present our results, discuss their implications in regards to attenuation of gas and dust or contamination in Section~\ref{sec:discussion}, followed by conclusions in Section~\ref{sec:conc}.

% * <cyril.simon.wedlund@gmail.com> 2017-03-30T22:39:00.686Z:
% 
% > we compare the photoelectron emission observed
% a lot of "we compare" in this sentence. Could be merged into one single sentence instead.
% 
% ^.

%{\color{red} Something here on what follows in the rest of the paper. In Section blabla we do blabla, und so weiter.}
%{\color{green} is this OK?//F}

% All papers should start with an Introduction section, which sets the work
% in context, cites relevant earlier studies in the field by \citet{Others2013},
% and describes the problem the authors aim to solve \citep[e.g.][]{Author2012}.

\section{Methods}
\label{sec:methods}

%The RPC-LAP instrument has several modes of operation to investigate the surrounding plasma environment, as described in detail by \citet{eriksson_rpc-lap:_2007}. Of particular interest is the Langmuir probe sweep (I-V curve), where we vary a bias potential from the spacecraft to the probe and measure the plasma current response at each step, with a typical cadence of once per minute. We can also set the bias potential to the probe constant and measure the current at much higher frequency

Conductive objects such as the Langmuir probes on Rosetta will emit electrons when subjected to sunlight due to the photoelectric effect. For cases when none of the electrons are reabsorbed by the probe, such as a negatively charged probe, the photosaturation current $I_{ph0}$ can be observed. The theory behind Langmuir Probe measurements is described in Section~\ref{sec:maths}, and three different techniques to observe $I_{ph0}$ is described in Section~\ref{sec:sunshadow}, \ref{sec:method1} and ~\ref{sec:method2}. We also propagate the UV observations around Earth and Mars to the Rosetta position, as described in Section~\ref{sec:propto67p}.

\subsection{Langmuir probe photoemission and probe theory}
\label{sec:maths} % used for referring to this section from elsewhere

A fundamental mode of operation of a Langmuir probe is the bias voltage sweep. During a sweep, the probe is measuring the current to the probe while stepping through a series of bias voltages $V_b$ set with respect to the spacecraft ground. The absolute potential $V_p$ between the probe and a plasma at infinity is thus $V_p = V_b + V_S$, where $V_S$ is the spacecraft potential. The current to the probe can be separated into three parts: ion ($I_i$), electron ($I_e$) and secondary electron emission current. The secondary emission current can be subsequently separated into photoemission ($I_{ph}$) and secondary electron emission from particle impact ($I_{SEEP}$).  
Akin to the photoemission current, $I_{SEEP}$ will depend on material properties of the probe but is also directly proportional to the ion and electron current to the probe such that if $I_e = I_i= 0$, $I_{SEEP} = 0$, and can in most cases be assumed to be negligible.

%{\color{red} Can photoemission be classified as secondary?}{\color{green} yes.the primary particle is in this case photons //F } 
% {\color{red} Suggestion: There are three main parts of the current to the probe: ion ($I_i$), electron ($I_e$), and electron emission currents. The electron emission current can be separated into photoemission ($I_{ph}$) and secondary electron emission due to particle impact ($I_{SEEP}$)}
% * <cyril.simon.wedlund@gmail.com> 2017-03-30T22:36:55.342Z:
% 
% > $I_{SEEP}$
% the use of capital letters here while all the rest is in lower case could maybe be improved for consistency. 
% 
% ^ <frejon@irfu.se> 2017-04-07T12:00:58.095Z:
%
% I understand what you're saying. But  I_seep doesn't look right to me, neither does I_E, I_I, etc. SEEP is an acronym with heritage. Like spacecraft heritage, it can't fail =)
%
% ^.

% Anders: https://en.wikipedia.org/wiki/Secondary_electrons

It can be shown \citep{medicus_theory_1961,mott-smith_theory_1926} that the electron current to a spherical probe is given by
\begin{equation}
 I_{e} =
  \begin{cases}
   I_{e0}\left(1+\frac{e V_{p}}{k_B T_{e}}\right) & \text{for } V_{p} \geq 0 \\
%   I_{e0}{e^{\frac{eV_{p}}{k_B T_{e}}}}    & \text{for } V_{p} < 0,
   I_{e0} \exp\left(\frac{eV_{p}}{k_B T_{e}}\right)    & \text{for } V_{p} < 0,

\end{cases}
  \label{eq:electron}
\end{equation}
where $I_{e0}$ is the random current for electrons, given by
\begin{equation}
  I_{e0} = A_p e n \sqrt{\frac{k_B T_{e}}{2 \pi m_{e}}},
\end{equation}
where $n$ is the plasma density, $T_e$ is the electron temperature and $A_p$ is the surface area of the Langmuir probe and other constants have their usual meaning. 
 %$e$ is the electron charge, and $k_B$ is Boltzmann's constant. %m_e missing... %engwall_cold_2006

For supersonic ion flow of single positive charge, the ion current to a sphere is shown by \citet{fahleson_ionospheric_1974} to be

\begin{equation}\label{eq:ion}
I_i = 
	\begin{cases}
	- I_{i0}\left(1 - \frac{ e V_{p}}{ E_i}\right) & \text{for } V_p < E_i/e \\
    0 & \text{for } V_p > E_i/e,
	\end{cases}
\end{equation}
where $E_i = \frac{m_i u^2}{2}$ is the energy of ions of mass $m_i$ and flow speed $u$, $I_{i0}$ is the ram current, given by
\begin{equation}
	I_{i0} = A_{c} e n u,
\end{equation}
where $A_c$ is the circular cross section of the probe.

For a negatively charged probe, the photoemission current is then at its saturation value $I_{ph0}$ which depends on the UV sun flux $F(\lambda)$ at each wavelength $\lambda$, the surface area normal to the sunlight $A_c$, as well as the photoelectron yield $PEY(\lambda)$ of the probe material, defined as number of emitted electrons per incident photon. Following \citet{grard_properties_1973}, and defining $F(\lambda)$ to be in units of photons~s$^{-1}$m$^{-2}$, we obtain:
\begin{equation}
I_{ph0} = - A_{c} \int{PEY(\lambda) F(\lambda) \, d\lambda} \, .
\label{eq:photoyield}
\end{equation}
%where $PEY(\lambda)$ is given in electrons per incoming photon, $F(\lambda)$, in photons $s^{-1}m^2$.

% * <cyril.simon.wedlund@gmail.com> 2017-03-30T22:35:35.204Z:
% 
% > \begin{equation}
% > I_{ph0} = - A_{c} \int{PEY(\lambda) F(\lambda) d\lambda}.
% > \label{eq:photoyield}
% > \end{equation}
% 
% how about using instead of the acronym PEY just $Y_\text{ph}$? In the studies of Grard and of Pedersen et al., what did they use as symbol for the yield?
% 
% ^ <frejon@irfu.se> 2017-04-06T23:54:51.562Z:
% 
% The popular (engineering) term is PEY, whereas the historical one is Quantum Yield (QY). There are those who write $Y_{e^-/\gamma}$, or W, but I'm hoping to avoid increasing the number of ways to write this. https://xkcd.com/927/ 
% 
% ^.

\begin{figure}
	\includegraphics[width=1.0\columnwidth]{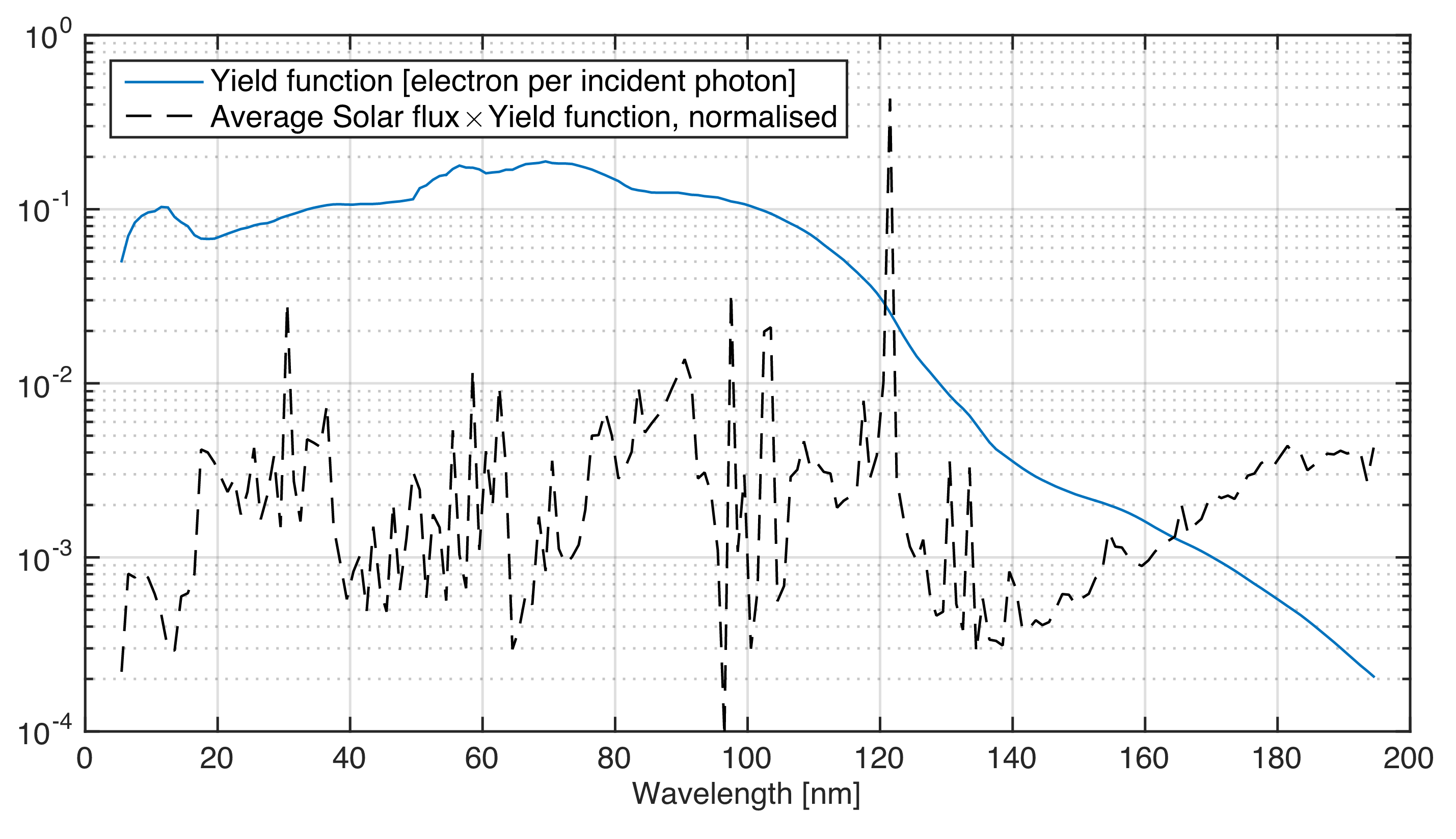}
    \caption{Photoelectric yield in electrons per incident photon (solid line) vs wavelength adapted from \citet{feuerbacher_experimental_1972,canfield_nbs_1987}, used to estimate photoemission from the probe from the two EUV datasets. The fractional current contribution of the typical value of each solar flux wavelength (dashed line) is also plotted for reference.}
    \label{fig:yields}
\end{figure}

In absence of photoelectron yield measurements for TiN, we follow the approach inspired by \citet{brace_solar_1988} and \citet{hoegy_how_1993} in a similar situation. They used a yield function adapted from an average of metals examined by \citet{feuerbacher_experimental_1972,canfield_nbs_1987} with a free numerical factor to scale the estimated photosaturation current from Equation~\ref{eq:photoyield} to the measured photoemission, using sun flux measurements of other spacecraft. In our case the free numerical factor turned out to be very close to 1 between the MAVEN/EUVM UV spectra and our photoelectron yield. We therefore adopted the yield function plotted in Figure~\ref{fig:yields} with no further tuning or correction.%we could therefore use the values obtained by Equation~\ref{eq:photoyield} without correction. The yield function used is plotted for reference in Figure~\ref{fig:yields}.

TIMED/SEE spectral irradiance data includes uncorrected degradation beginning in late 2011, resulting in irradiances that become increasingly lower with time than those measured by other spacecraft (SDO/EVE). Therefore the free numerical factor differed from 1 when we used TIMED/SEE data.

\subsection{Probe photoemission current from Sun-shadow transitions}
\label{sec:sunshadow}
The component of the total probe current due to photoemission from the probe surface, $I_{ph}$, clearly vanishes for a probe that is not sunlit. Therefore, an obvious way of measuring the photoemission current of a probe is by comparing the probe current at fixed voltage-bias just before and after it goes into or out of shadow. This seemingly straightforward method requires at least two conditions to be met: (1) the probe must be at a negative potential w.r.t.\ the local plasma at the position of the probe in order for the full photosaturation current $I_{ph0}$ to be sampled, limiting the dataset to probe sun/shade transitions during which LAP1 is commanded at a negative bias potential w.r.t.\ the spacecraft; (2) concurrent variations in probe current due to other factors, e.g.\ varying plasma density, temperature, drift velocity etc., must be either negligible in comparison to the photoemission current, or occur on sufficiently short time-scales that their effects can be filtered out. In practice, this limits the applicability of the method to probe sun/shade transitions that are sufficiently fast for the general background ion current to be essentially a stationary process, but at the same time separated by sufficient time for calculation of statistical moments of this process, e.g.\ arithmetic mean and standard deviation. We have in this study decided to use only probe sun/shade transitions in which the probe goes from completely sunlit to completely shaded, or vice versa, in no more than two minutes, preceded and succeeded by periods of complete sunlight or shade for at least two minutes.

\begin{figure}
	\includegraphics[width=1.0\columnwidth]{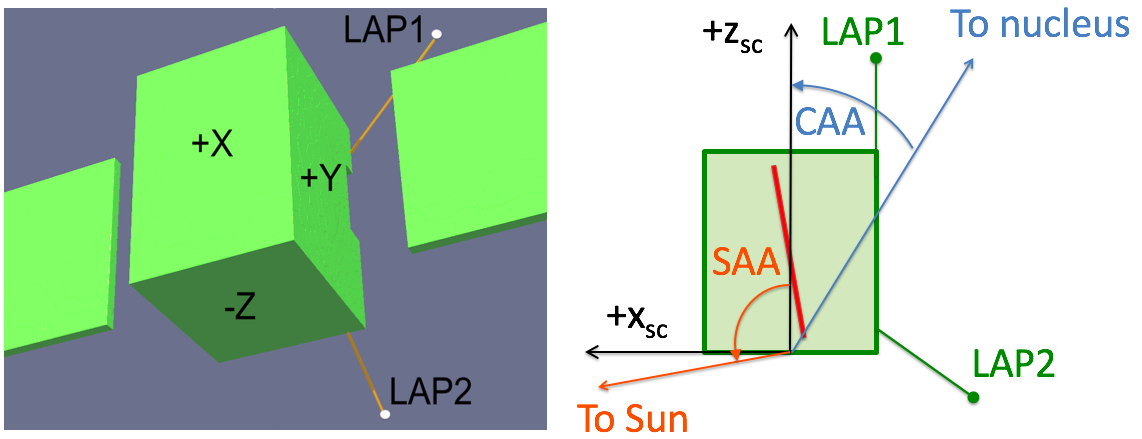}
    \caption{\textbf{Left:} 3D-visualisation of the Rosetta spacecraft with the two Langmuir Probes LAP1 and LAP2. \textbf{Right:} Geometry visualisation and definition of Solar Aspect Angle (SAA) and Comet Aspect Angle (CAA), solar panels marked in red. When the spacecraft box turns around its Y-axis, LAP1 goes in and out of shadow behind the +Y Solar array for a certain range of SAA.}
    \label{fig:viz}
\end{figure}

Figure~\ref{fig:viz} (left) shows a sketch of the Rosetta spacecraft and RPC-LAP. The solar panels were almost always held orthogonal to the Sun, meaning the S/C Y-axis stayed perpendicular to the sun. When the spacecraft turned around its Y-axis, which happened regularly, LAP1 would become completely shadowed by the spacecraft solar array. In Figure~\ref{fig:viz} (right) we define the Solar Aspect Angle (SAA), as the angle between the spacecraft +Z and the direction of the Sun, counted positive when the Sun moved from +Z toward +X.

\begin{figure}
	% To include a figure from a file named example.*
	% Allowable file formats are eps or ps if compiling using latex
	% or pdf, png, jpg if compiling using pdflatex
	\includegraphics[width=1.0\columnwidth]{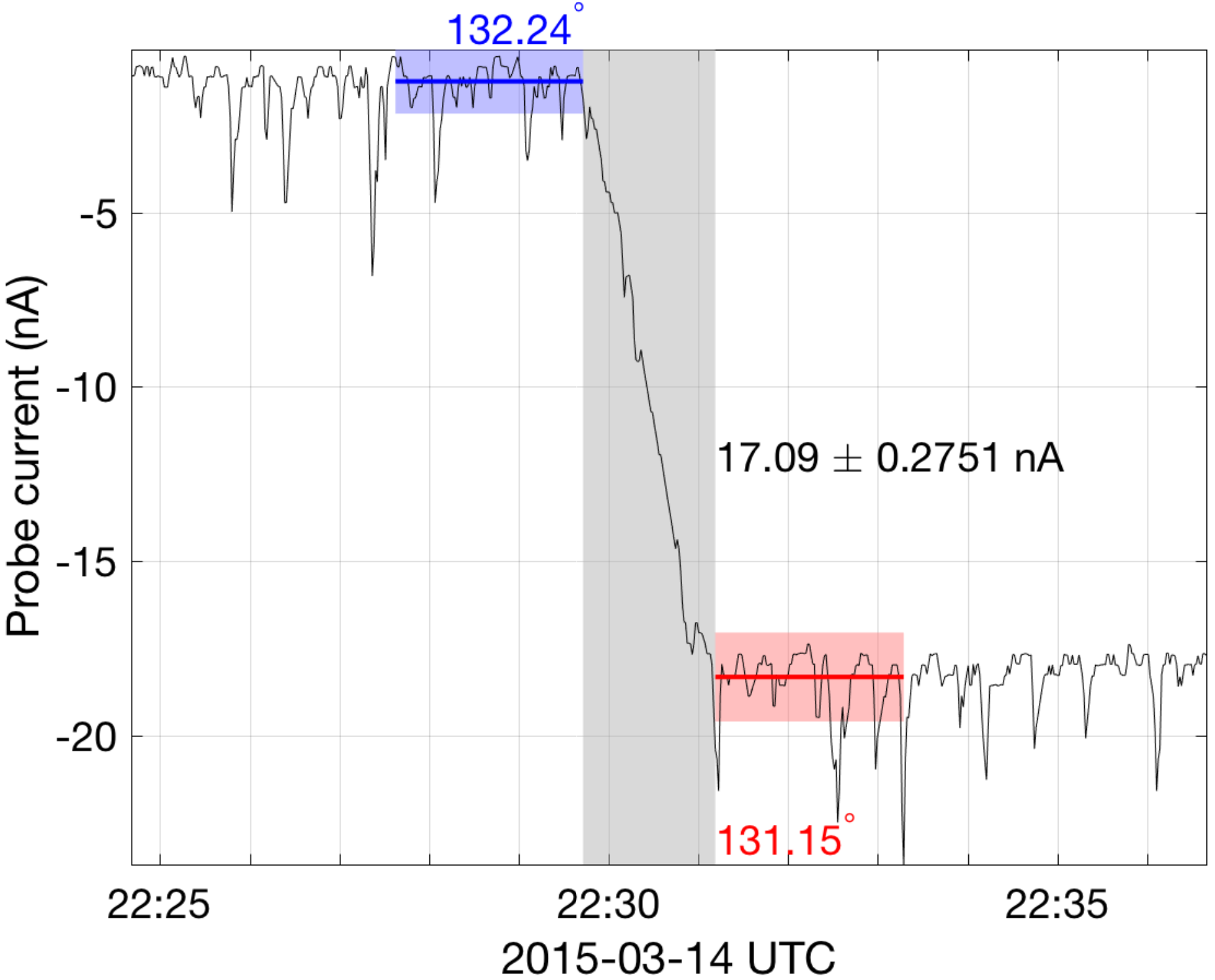}
    \caption{Example of shadow-crossing $I_{ph0}$ estimate from current level shift when crossing into shadow for a negatively charged probe.}
    \label{fig:iph0_shadow}
\end{figure}

Figure~\ref{fig:iph0_shadow} shows an example from March 3, 2015, where LAP1 goes from shadow to sunlight. The actual sun/shade transition, during which the probe is partially sunlit and hence draws a successively increasing photoemission current, is marked by the grey patch in the figure. The solar aspect angles at which the probe enters and exits partial illumination conditions are shown in the figure; they are $132.2^\circ$ and $131.2^\circ$, respectively. For LAP1 there is also a second possible transition region between $178.2^\circ$ and $179.2^\circ$ which is also used in this study, when available. For visual reference, see Figure~\ref{fig:viz}.

The magnitude of the current jump across the transition region is calculated by taking the difference of the arithmetic means of the probe current during the 2-minute periods immediately preceding and succeeding it, shown as blue and red lines, respectively, in Figure~\ref{fig:iph0_shadow}. As can be seen in Figure~\ref{fig:iph0_shadow}, the probe current is prone to brief pulses of considerably increased magnitude. These have been interpreted as cold plasma filaments passing by the spacecraft \citep{eriksson_cold_2017} and produce a substantially skewed distribution about the mean of the sample currents. Therefore, following \citet{tukey_exploratory_1977}, all sample currents that lie more than 1.5 times the interquartile range below the first quantile or above the third quantile are discarded as outliers.

Sample standard deviations $\sigma_{\textrm{before}}$ and $\sigma_{\textrm{after}}$ are computed taking into account sample auto-correlation using the method of \citet{zieba_effective_2010} (specifically their Equation (10)). Corresponding confidence intervals at the 95\% level, $1.96\sigma/\sqrt{n}$, with $n$ the number of samples, are shown in Figure~\ref{fig:iph0_shadow} as blue and red shaded regions, respectively. A confidence interval for the difference of the means is then simply obtained as $1.96\sqrt{\sigma_{\textrm{before}}^2/n_{\textrm{before}} + \sigma_{\textrm{after}}^2/n_{\textrm{after}}}$.

This method for measuring the probe photoemission current has the advantage of being unaffected by any potential additive offsets in the current measurements, since it relies on a current difference and not the absolute value. Its main shortcoming is that its application is contingent on specific attitude and commanding criteria, as previously mentioned, and as thus a rather sparse and uneven dataset. During the entire autumn of 2014, there is not a single sun/shade transition of LAP1 coinciding with commanded negative fixed bias voltage. This situation was somewhat remedied by the fact that the ion current in the frequent sweeps during this time of low cometary activity far away from the Sun was entirely negligible compared to the photoemission current. Therefore, photoemission estimates from August to October 2014 was obtained from the difference of the currents at large negative bias voltages between sweeps immediately before and after a sun/shade transition. Specifically, we obtain arithmetic means and standard deviations from the currents at the lowest 5 V of bias potentials in each sweep, allowing us to estimate confidence intervals of the photoemission current for each transition.

% This method for measuring the probe photoemission current has the advantage of being unaffected by any potential additive offsets in the current measurements, since it relies on a current difference and not the absolute value. Also, it is completely independent of any possible shortcomings of the sweep analysis which could hypothetically affect the other methods used in this Paper. It's main shortcoming is that it's application is contingent on specific criteria on the spacecraft attitude, viz.\ the solar aspect angle, which are only very intermittently and irregularly fulfilled during the mission, leading to somewhat uneven coverage of the resulting photoemission estimates. E.g.\ during the entire fall of 2014, there is not a single sun/shade transition of LAP1 coinciding with negative fixed bias voltage. This situation can be somewhat remedied by the fact that the ion current in the sweeps during this time of long cometary activity far away from the Sun was entirely negligible compared to the photoemission current. Therefore, photoemission estimates from this time period can be obtained from the difference of the currents at large negative bias voltages between sweeps immediately before and after a sun/shade transition. Specifically, we obtain arithmetic means and standard deviations from the currents at the lowest 5 V of bias potentials in each sweep, allowing us to produce confidence intervals of the photoemission current for each transition.

\subsection{Probe photoemission from single sweeps} \label{sec:method1}

\begin{figure}
	\centering
	\includegraphics[width=0.75\columnwidth]{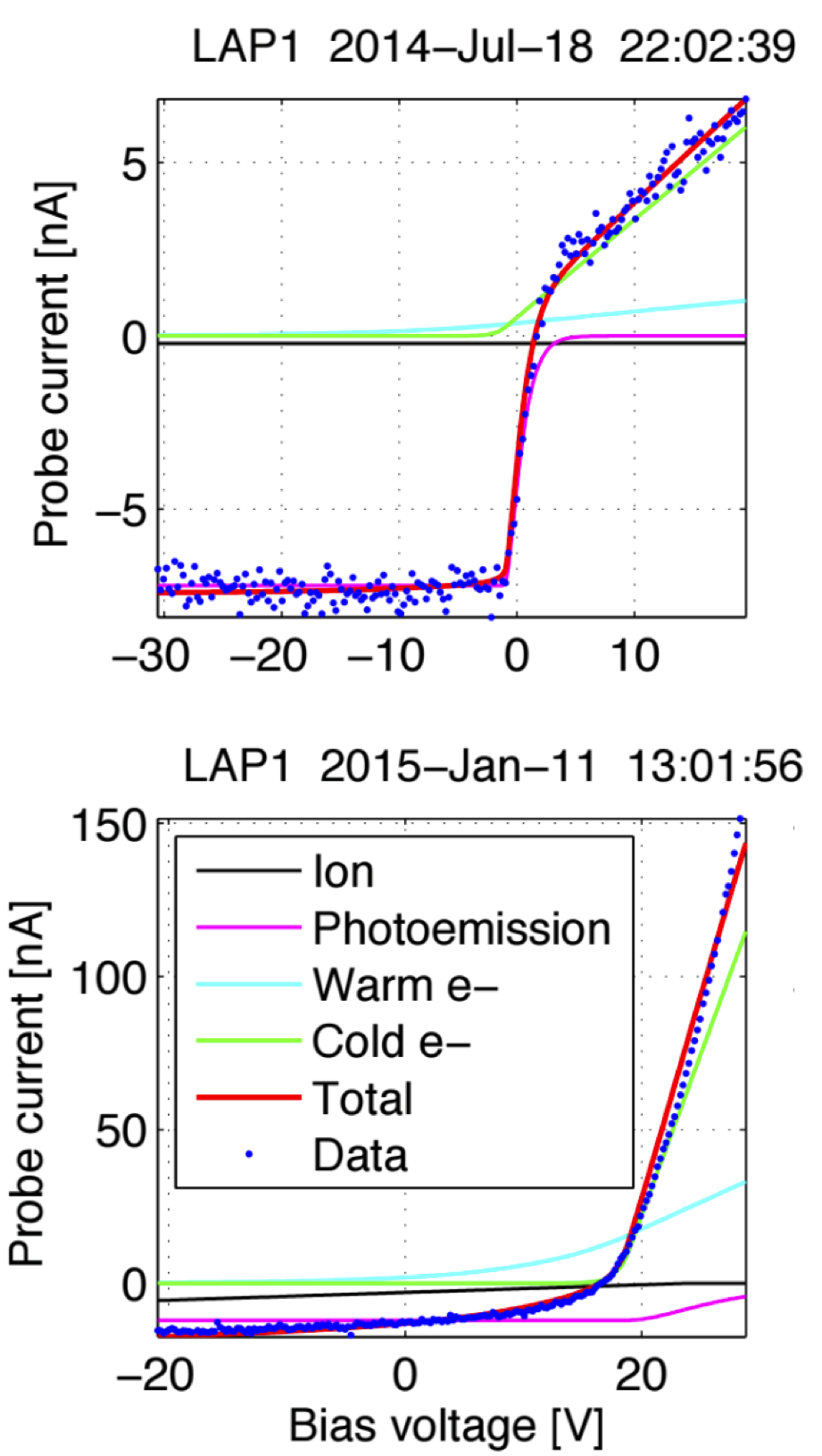}
    \caption{Example RPCLAP I-V curve data (blue) and photoemission (pink) model for two different plasma regions of tenuous (\textbf{Top}) and dense (\textbf{bottom}) plasma, figure courtesy of \citet{eriksson_cold_2017}. %{\color{green}todo : Make a new plot  //F}
}
    \label{fig:example_IVcurves}
\end{figure}

Throughout the Rosetta mission, the Langmuir probe instrument has seen a very dynamic and varying plasma with regions where ion, electron and photoemission current have, within the bias voltage range of the Rosetta Langmuir probe, each individually dominated the Langmuir probe sweep measurement \citep{eriksson_cold_2017}. For sunlit probes, an automatic routine was set to find the knee $V_{ph}$ in the I-V curve, as previously described by \citet{odelstad_measurements_2016}, and by proxy, $V_S$, to subdivide the I-V curve into two regions of $V_p$. Assuming the spacecraft potential is well within the voltage bias sweep of $\pm~30$~V, the two regions will be characterized by the linear ion current and the photosaturation current for $V_p <~0 $ and a linear electron current for $V_p > 0$. Using the fact that at $I(V_p = 0) \approx I_{e0} + I_{i0}$ 
from Equations ~\ref{eq:electron} and \ref{eq:ion}, we use an automatic fitting routine to remove a model of the electron and ion current component to obtain only the current contribution from $I_{ph}$ and $I_{SEEP}$. The latter can be assumed to be negligible for all but the densest of plasmas, such that for negative $V_p$ we can obtain $I_{ph0}$. An example of the sweep current fitting routine result is shown in Figure~\ref{fig:example_IVcurves}.

%, as well as the noise level of the instrument%
% Anders kommenterade att brusnivån borde vara låg i jämförelse, men inte för >3AU, och möjliga MIP störningar. Jag känner att det kanske är värt att hålla kvar den meningen.

The accumulated errors from a single estimate with this technique is expected to be large, owing to the many mutually dependent fits needed to procure the estimate, as well as the noise level of the instrument. Indeed, there are times were the automatic routine does not produce physically meaningful results, and as such only 92\% of the dataset was used due to inexactness of the automatic routine, interference or erroneous commanding. The automatic routine and the threshold of validity used are still in development, so to limit the impact of erroneous estimates, we present the median result over an operational block, defined as a period when the instrument is operated in one single operational mode (on average 400 measurements over 3-4~hours), and the Median Absolute Deviation (MAD) in Figure~\ref{fig:iph0fig1}.

%a continuous mode of operation (on average 400 measurements over 3-4~hours) and the Median Absolute Deviation (MAD) in Figure~\ref{fig:iph0fig1}.

The largest source of random error is estimated to arise in the electron model fit, in a region where we often see a non-well-behaving current, as well as the difficulties of correctly estimating $V_S$, as studied in greater detail by \citet{odelstad_measurements_2016}. However, as the sample size is very large (around 400,000) we expect to become much less sensitive to random errors as we take a median of the results. Of all uncertainties involved, the dominant source is expected to be the possible systematic error from secondary emission current. which may give exaggerated values of $I_{ph0}$. If there is a discrepancy between this result and other $I_{ph0}$ estimates, we may be able to estimate the impact of secondary electron emission from particle impact on the Langmuir Probes.

%Therefore, the sources of systematic errors may be more important, for which we have considered two: $I_{SEEP}$, which we expect to offset our estimation of $I_{ph0}$ to larger values for dense plasma, and instrumental offsets still present after calibration.
%{\it Fredrik}

\subsection{Probe photoemission from analysis of multiple sweeps }
\label{sec:method2}
%{\it Joakim, Sigve -- probably pretty close to done}

Assuming that the photoemission from the probe does not change significantly between sweeps, it is possible to find the photoemission current by combining results from several sweeps. For this purpose, data from the ion saturation region is required, so that the electron current can be taken as negligible. Taking $eV_p/k_B T_e \ll -1$, the total current, assuming no secondary emission by particle impact, detailed in section~\ref{sec:maths} simply reduces to
\begin{equation}\label{eq:Itot_sec24}
I_{tot} \approx I_i + I_{ph0} ,
\end{equation}
and the derivative is then
\[
\frac{d I_{tot}}{d V_p} = \frac{d I_{i}}{d V_p}.
\]
This result combined with Equations~\ref{eq:ion} and~\ref{eq:Itot_sec24} yields
\begin{equation}\label{eq:slope_v_current}
%\frac{d I_{tot}}{d V_p} = k \left( I_{tot} - I_{ph0}\right) = \frac{2 A_c n e}{m_i u},
\frac{d I_{tot}}{d V_p} = k \left( I_{tot} - I_{ph0}\right) \propto \frac{n}{u},
\end{equation}
where $k$ is given by
\[
k = \frac{e}{e V_p - E_i}.
\]

Assuming that $k$ does not change during the sweeps, we can extrapolate for $I_{tot}(n=0)$ from a number of measurement of $I_{tot}$ and $\frac{d I_{tot}}{d V_p} $ with enough spread in $n$, as

\begin{equation}
  \frac{d I_{tot}}{d V_p}(n=0) = 0 \Rightarrow I_{tot} \approx I_{ph0}.
  \label{eq:didv}
\end{equation}

%By assuming that $k$ do not change during the sweeps, it is possible to combine them and extrapolate the data to find the photoemission current using Equation~\ref{eq:slope_v_current} and the fact that
%\begin{equation}
%  \frac{d I_{tot}}{d V_p} = 0 \Rightarrow I_{tot} \approx I_{ph0}.
%  \label{eq:didv}
%\end{equation}

Although we have so far ignored $I_{SEEP}$, we note that for electron emission from ion impact, if it increases linearly with density and for $eV_p \ll E_i$ it would be indistinguishable from the ion current and would not affect Equation~\ref{eq:didv}. For other values of $E_i$ it would introduce a small non-linear current slope as the energy of primary collision species increases. Also, as the primary electron current in the specified region is assumed to be negligble, so would its secondary current be. When this is not the case, the secondary current would instead slightly mitigate the effect the primary electron current would have on the photoemission estimate.

%{\color{green} yes, this is correct. The only error source that would show is electron impact emission. (although this is not the case for method 1)}
%For several sweeps at different plasma densities a linear fit of slope of the ion current and the current at an arbitrary voltage $V_p < 0$ will yield a solution at $I_{tot} = I_{ph0}$ when the fit crosses the x-axis.
\begin{figure}
	% To include a figure from a file named example.*
	% Allowable file formats are eps or ps if compiling using latex
	% or pdf, png, jpg if compiling using pdflatex
	\includegraphics[width=1.0\columnwidth]{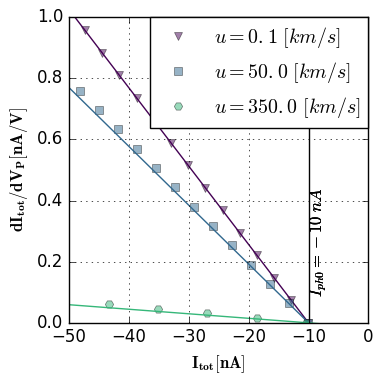}
    \caption{Slope versus $I_{tot}$. For several sweeps at different plasma densities, a linear fit of the ion current and slope will yield an estimate of $I_{ph0}$ when the fit crosses the x-axis. Triangle, square, and circle points are taken from synthetic data sweeps with $T_e = 5$~eV, $V_S =$-10~V and varying $n$ for three different ion velocities, and the solid lines are calculated from Equation~\ref{eq:slope_v_current}. There is a good agreement between the synthetic data and Equation~\ref{eq:slope_v_current}. Small deviations are related to high plasma densities, where $I_e$ becomes significant.}
    \label{fig:error_theory}
\end{figure}

Restricted by the condition that $k = e/(e V_p - E_i)$ needs to be approximately constant during the series of sweeps, it is worth investigating under which range of plasma parameters this method is effective. Note that in order to use Equation~\ref{eq:slope_v_current} to extrapolate $I_{ph0}$ we need several data points from sweeps during changing plasma conditions, as illustrated in Figure~\ref{fig:error_theory}. Although this method is sensitive to both $n$ and $u$, the plasma environment around Rosetta varies much faster in density, with order of magnitude density fluctuation time-scales of minutes to seconds as reported by \citet{henri_localised_2016} as well as the diurnal variation evident by the $V_S$ data in \citet{odelstad_evolution_2015}. In comparison, the ion velocity appears much more stable (Vigren et al 2017, this issue), but will still introduce some random error. We investigate this theoretically in Appendix~\ref{sec:app_oslo}. In Section~\ref{sec:obs} we will find that all three methods agree well, suggesting small errors in practise.

The $I_{ph0}$ dataset from the multiple sweep method and the variance of the fit is plotted in Figure~\ref{fig:iph0fig1}, Results with large variances in the linear ion slope, and as such, large non-linear effects from e.g. the electron retardation current, are discarded.

%and is assumed to be constant over a few hours in this study. 

%Ideally the ion energy would remain constant and only the plasma density changes between the sweeps, creating a set of data points that, within the noise level, perfectly line up (still assuming negligible $I_e$) and Equation~\ref{eq:slope_v_current} would give an accurate estimate of the photoemission  current, $I_{ph0}$. However, since the ion energy does change between the sweeps, this may lead to different $k$ and consequently, introduce inaccuracy in the estimation of $I_{ph0}$.

\subsection{Propagation of TIMED/SEE and MAVEN/EUVM data to Rosetta}
\label{sec:propto67p}

Rosetta does not carry any instrument for direct measurement of solar UV flux at the position of the spacecraft. For comparison, we rely on measurements from TIMED/SEE at Earth and MAVEN/EUVM at Mars, propagated out to the position of 67P. The propagation procedure consists of finding the last preceding and first succeeding epochs at which the 'source', Earth or Mars, respectively, was at the same solar longitude as that of Rosetta at the queried epoch. The measured UV flux at these epochs are then scaled by the square of the respective heliocentric distances of the source divided by that of Rosetta at the queried epoch. Finally, a weighted average of the two scaled EUV measurements at the source is computed such that the value of the closest measurement taken less than a few days before or after the queried epoch was used, otherwise the two values are weighted together by a linear interpolation of their respective temporal separation from the queried epoch.

%Finally, a weighted average of the two scaled EUV measurements at the source is computed such that the value of the closest measurement taken less than 7 days before or after the queried epoch was used, otherwise the two values are weighted together by the inverse of their respective temporal separation ($<7$~days) from the queried epoch.

%that if the closest measurement was taken less than 7 days before or after the queried epoch that value is used, otherwise the two values are weighted together by the inverse of their respective temporal separation (less 7 days) from the queried epoch.

The datasets used for the propagation are Level 3 daily averages from the TIMED/SEE database \citep{woods_solar_2005} and Level 3 daily averaged spectral irradiance (non-flare background data) from MAVEN/EUVM \citep{thiemann_maven_2017} and the propogated results are plotted in Figure~\ref{fig:iph0fig1}. The shaded regions in Figure~\ref{fig:iph0fig1}a corresponds to periods where Mars (purple) and Earth (green) was within a 45$^\circ$ azimuth sector of Rosetta in the elliptical plane, where we expect optimal correlation. It should be noted that Mars never was further than 72$^\circ$ behind Rosetta until July 2016.

\section{Observations}
\label{sec:obs}

%$\text{I}_\text{ph0}$ 
\begin{figure*}
	\includegraphics[width=2.0\columnwidth]{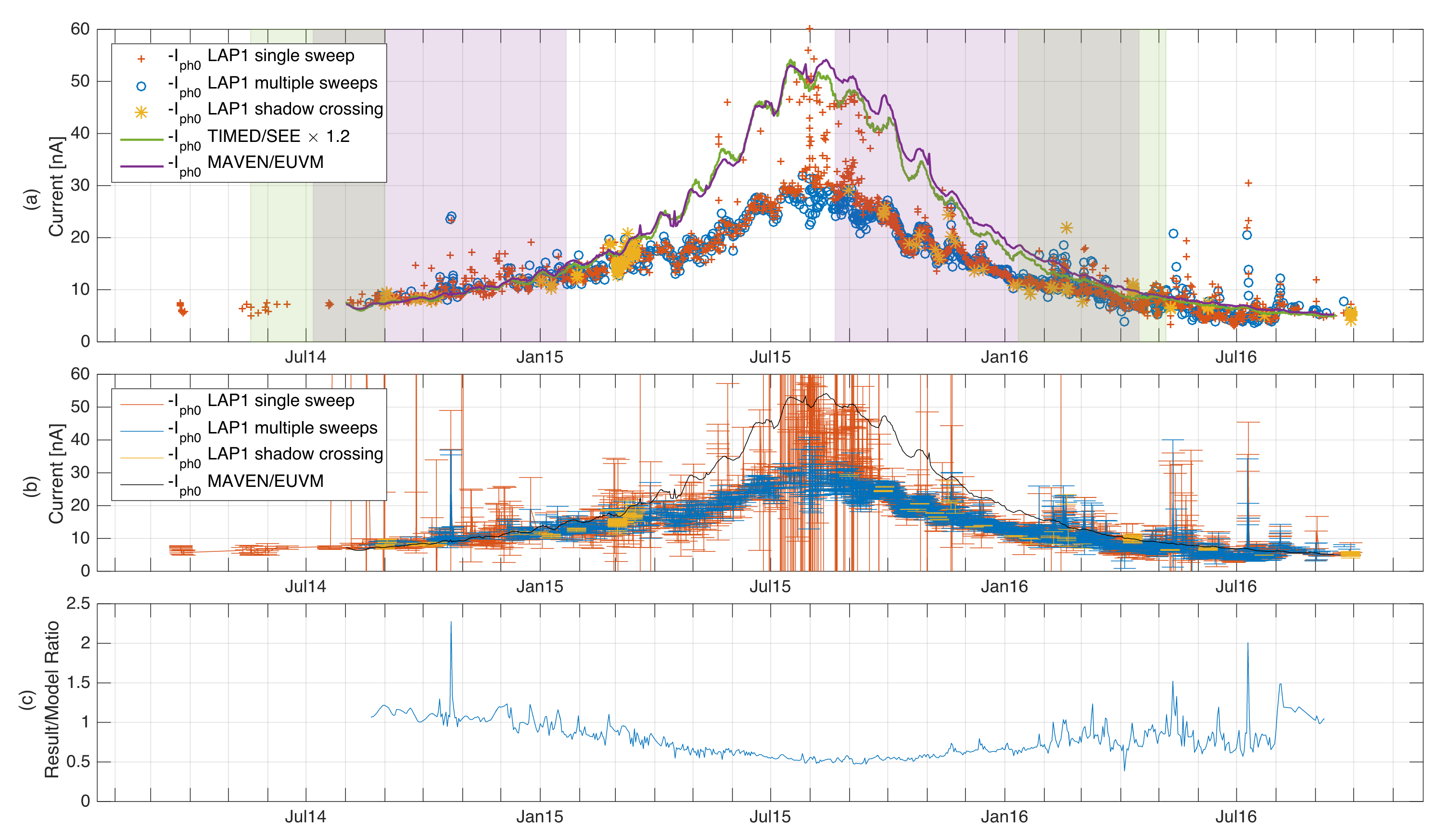}
    \caption{\textbf{Top:} Photo-saturation current and estimated photoemission current from EUV data from March 2014 to September 2016. The shaded regions corresponds to periods where Mars (purple) and Earth (green) was within a 45$^\circ$ azimuth sector of Rosetta in the elliptical plane, where we expect optimal correlation. The $I_{ph0}$ estimate from TIMED/SEE EUV data was scaled to align with the MAVEN/EUVM dataset with a factor of 1.2. \textbf{Middle:} Result and variance of the three different methods used, and the MAVEN/EUVM $I_{ph0}$ estimate for reference in black. Periods of large variance coincide with solar flares for the multiple sweep method and both flares and comet activity for the single sweep method. \textbf{Bottom:} Ratio between LAP1 multiple sweep method $I_{ph0}$ result and $I_{ph0}$ MAVEN/EUVM model, interpolated to one estimate per day.}
    \label{fig:iph0fig1}
\end{figure*}
%The normalised product of a typical solar EUV flux and the yield function (dashed line) is also plotted for reference}
%{\it Fredrik, all}

All three methods to obtain the photo-saturation current correlate well with each other both on the global scale and the small scales in Fig~\ref{fig:iph0fig1}a, resolving the solar sidereal rotation period of $\approx$~24.5~days. In particular, the good agreement between the three different methods verifies the methods and suggests that the we are not heavily influenced by the many individually unique error sources for each method.
%This verifies the methods, and an indication that the many unique error sources for each method does not have a large impact on the result. 
%In particular, the good agreement between the three different methods suggest 

The two EUV photoemission estimates derived from MAVEN/EUVM and TIMED/SEE data agree over the entire period with negligible differences between them if we scale the TIMED/SEE data by a factor of 1.2, in an attempt to correct for the known degradation on TIMED/SEE since late 2011. The scaling could also be analogous of a slightly larger photoelectron yield, and is still very reasonable from the photoelectron yield dataset \citep{feuerbacher_experimental_1972}.

The photosaturation current obtained from single sweeps cannot distinguish between secondary emission from photon or particle impact, and around perihelion (Aug 2015), where we expect high densities and collisions to be more frequent, we obtain a much higher estimate than the other methods at certain points. As evident by the large variance of these points in the single sweep method in Figure~\ref{fig:iph0fig1}b, the uncertainties at perihelion are large, and as such, these points of the single sweep method data should be ignored. However, subtracting the single sweep $I_{ph0}$ estimates from the multiple sweep $I_{ph0}$ estimates, we could obtain estimates of the secondary emission from particle impact from the probe (although not done in this report). This is otherwise impossible to observe directly with LAP. We can also use this to estimate when we safely can assume impact emission to be negligible.

%\subsection*{Flares}

Also evident in the result and variance of the $I_{ph0}$ data are brief large-amplitude changes of photoemission, which upon inspection correlate with solar flares (e.g. X1.6 flare 2014-10-22, M4.5 flare 2015-11-20).
%\citep{gibeau_solarham.com_nodate}).

The mission typical 30-160~s cadence of the Langmuir Probe sweep is in theory more than enough for detecting most flares in EUV \citep{aschwanden_stereo/_2014,thiemann_time_2017}. However individual probe sweeps may have significant noise, due to the plasma conditions and other error sources, complicating such detection. More than anything the processing (binning, averaging) of the dataset in this report limits the detection to Long Decay Events (LDEs) with durations on the order of hours, typically reserved for the largest GOES X-ray flare classes (X and M) \citep{aschwanden_automated_2012}. In a study by \citet{aschwanden_stereo/_2014}, they conclude that their channel of largest wavelength, 30.4~nm is the optimal for detecting flares in the EUV. As seen in Figure~\ref{fig:yields}, the RPC-LAP probes can be expected to have good sensitivity to these wavelengths.% which does not make an insignificant contribution to the LAP current as seen in Figure~\ref{fig:yields}.

%Similarly to flare detectors in EUV \citet{aschwanden_stereo/_2014}, the cadency of the Langmuir Probe sweeps is more than enough to detect flares

%Due to the impulsive nature of flares, the time resolution of the sweeps (Typically 1-3 second measurements every 30-160 second) and bias current measurements detects the number of flare we can detect. 

%Detection of flares in the $I_{ph0}$ data is interesting. The best chance to detect it is in the single sweep analysis, since the shadow transition method is either very infrequent, and the analysis in the multiple sweep method is done in such a way that a very short flare would be averaged away, but perhaps detectable in an increased spread.

%The algorithm for detecting flares is not finalised, but we offer no other explanation for signals that are more than 1$\sigma$ above the two EUV estimates, which are both non-flare background data products \citep{thiemann_time_2017,woods_solar_2005}. We also find correlation with MAVEN/EUVM 17-22nm flare signals below 1$\sigma$ and possibly also a few false detections, and needs to be studied further.

% * <cyril.simon.wedlund@gmail.com> 2017-03-30T22:52:17.093Z:
% 
% add reference here to Edberg et al. 2016. Regarding flares, maybe we should say that only large flares (M and X), that usually last in the 15 min-1 hour range are clearly seen in the LAP data. Since it depends on the sweeps technique you use, you even have a minimum length of flares that could be derived from the resolution of the LAP data.
% 
% ^.

\section{Discussion}
\label{sec:discussion}

As can be seen in Figure~\ref{fig:iph0fig1}c, the UV flux derived from probe photoemission is about as expected at the start and end of the mission, but smoothly drops to about half its expected value around perihelion. There seems to be some shortfall also at end of mission, but data variability here is much more pronounced. Nevertheless, it is clear that the photoemission has a drop around perihelion, from which it at least partially recovers as Rosetta follows the comet outward. 
%A surprising result was that the long term evolution shows less photoemission than expected around perihelion.  

We will discuss possible explanations of this photoemission decrease including contamination in Section~\ref{sec:disc_contamination} and attenuation by comet gas or dust in Section~\ref{sec:disc_gas} and \ref{sec:disc_dust}, but we note first that: (1) the orbital inclination of Comet 67P of $7^{\circ}$ from the planetary ecliptic plane, would give us a slightly different Sun flux than measured by MAVEN/EUVM or TIMED/SEE, but the effect would be limited and much less than observed; (2) the good agreement for all three methods suggests that we are not significantly affected by offsets and unique individual error sources; (3) the uncertainty in the photoelectric yield of TiN may affect the estimated $I_{ph0}$ on small scales, but even a vastly different yield profile would not change the deviation around perihelion since there is no significant trend in any wavelength channel that influences the result other than the $r^{-2}$ dependence over the course of the mission.

%Taking the uncertainties of the TiN yield profile aside, even a vastly different yield profile would not change the deviation around perihelion since there is no significant trend in any wavelength channel that influences the result more than the $r^{-2}$ dependence.

%The overall fit to the EUV estimates is excellent on small scales and at large heliocentric distances, but diverges at perihelion. Assuming we have a clear estimate of the solar EUV before encountering the comet in early 2014, we appear to be missing photoemission current at perihelion, and the authors have yet to find a simple explanation for this. 
% * <cyril.simon.wedlund@gmail.com> 2017-03-30T22:42:58.810Z:
% 
% > great
% "good" is enough here. "Excellent" if you want to push it! 
% 
% ^ <frejon@irfu.se> 2017-03-31T12:43:02.527Z.

%Comet 67P has an orbital inclination of $7^{\circ}$ from the planetary ecliptic plane, and would thus see a slightly different Sun flux than captured by both MAVEN/EUVM or TIMED/SEE at perihelion. However, the effect of this would be limited and much less than observed.

\subsection{Contamination}
\label{sec:disc_contamination}

Effects on the probe surface such as contamination, which could introduce a resistance and a net reduction in emitted current, would either be expected to be cumulative over the entire mission or be less during periods when the comet-spacecraft distance is large such as the day-side (Sep-Oct 2015) or night-side excursion (Mar-Apr 2016). In addition, a contamination in form of a resistive and capacitive layer should be discernible when alternating bias stepping direction (hysteresis sweeps) according to \citet{szuszczewicz_surface_1975}:

\[
\Delta I = \frac{C\Delta V_b}{\Delta t}
\]
% \begin{equation}
% %C = \frac{Q}{\Delta V_b},
% \Delta I = \frac{C\Delta V_b}{\Delta T}
% \label{eq:Contam}
% \end{equation}
where $\Delta t$ is the time between two subsequent current measurements $\Delta I$ on a probe with a capacitance C.

%{\color{green} I'll probably move this following paragraph to the appendix  //F},
%The full dataset of hysteresis sweeps are discussed in Appendix~\ref{sec:app_cont}, and shows only 

\begin{figure}
	% To include a figure from a file named example.*
	% Allowable file formats are eps or ps if compiling using latex
	% or pdf, png, jpg if compiling using pdflatex
   
   \centering 
	\includegraphics[width=0.9\columnwidth]{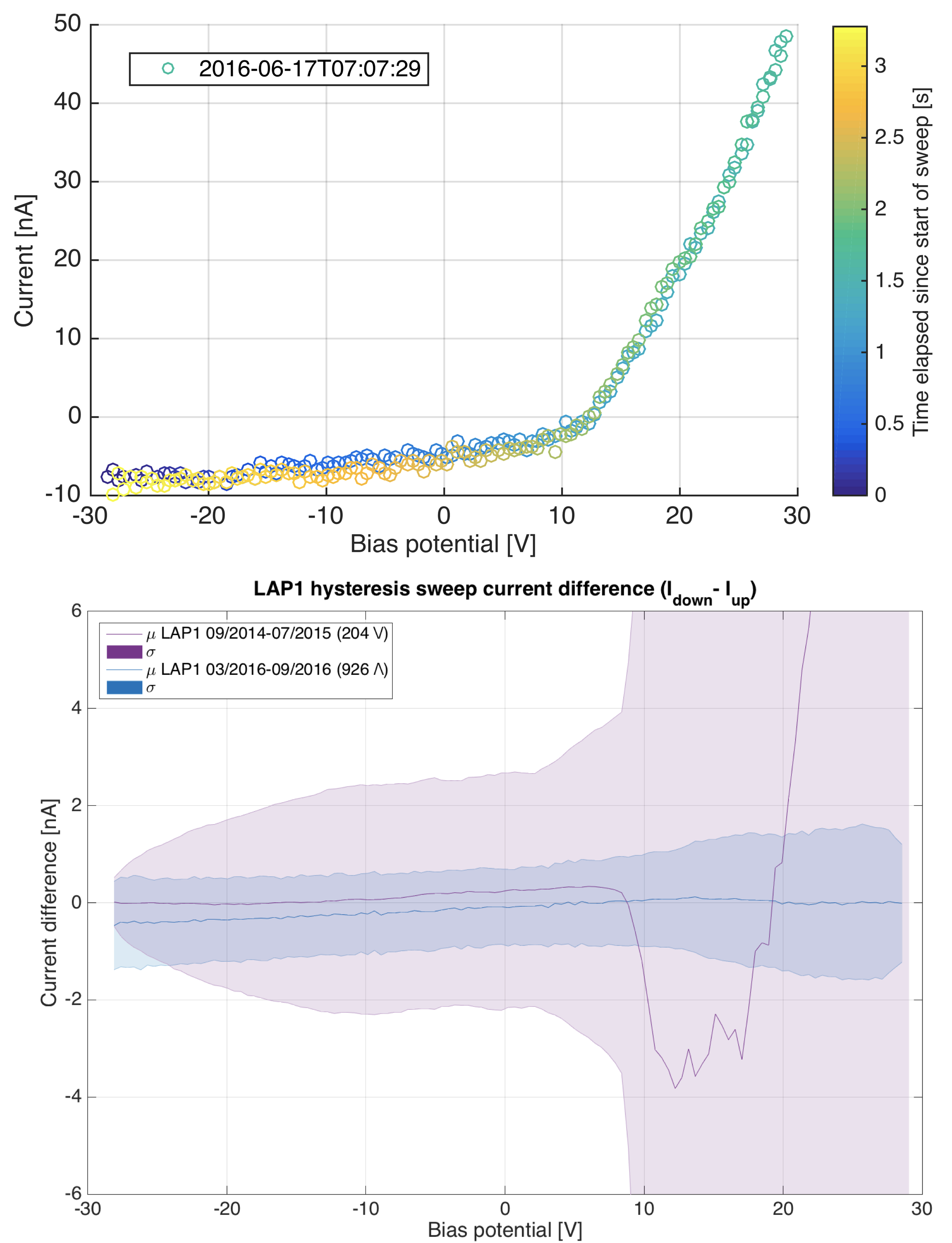}
    \caption{\textbf{Top:} Example LAP1 hysteresis up-down sweep from blue to green to yellow. \textbf{Bottom:}Hysteresis sweep analysis during the comet phase of the mission divided into two datasets with different sweep parameters. Averages and standard deviation  of down-up sweeps during 2014-2015 in purple, and up-down sweeps (2016 data) in blue.}
    \label{fig:hyst}
\end{figure}

To monitor contamination on the Langmuir probes, more than 23 000 hysteresis sweeps were performed throughout the mission. A summary for LAP1 is plotted in Figure~\ref{fig:hyst}, subdivided into two datasets with different starting potentials and time periods.  The sweeps are either from -30~V up to + 30~V and back down to -30~V ('up-down') or vice versa ('down-up'). %The high variance (up to 400~nA, off scale) in the down-up sweeps at positive potentials is believed to be due to plasma changes being more noticeable in the electron current when the elapsed time between the measurements is large.

As evident by the large variance (up to 400~nA, off scale) for the first set of sweeps, the electron current to the probe changes rapidly in comparison to the sweep duration ($\approx 6$~s), such that the method was changed during 2016 to an up-down type of sweeps, with clearer results. We find no significant capacitive current contribution, but estimate that at most it would offset our results in the two sweep analysis methods with $0.3(\pm 0.5)$~nA, and is as such negligible. The sun-shadow transition data are unaffected by this capacitive current offset.

On LAP2, we do find evidence of significant contamination, particularly during a few months after the day-time excursion in Oct 2015 with a capacitive current contribution exceeding 20~nA, as well as a significant decrease in photoemission current. The LAP2 results are therefore excluded from this report.

\subsection{Attenuation by comet gas} \label{sec:disc_gas}

As observed by the Rosetta Alice instrument \citep{keeney_h2o_2017}, the neutral gas and dust of the comet coma can absorb a large ($>50$ percent) fraction of some spectral lines in the 70-200 nm range of Alice observation. In particular, the common comet gas species of interest with substantial absorption cross-sections within our yield profile would be $\text{H}_2\text{O}$, $\text{CO}_2$ and $\text{CO}$. However, Alice absorption observations are along a path close to the comet core, with regions of peak densities, whereas the Probe-Sun path are almost always along more tenuous atmosphere profile due to the terminator plane or dayside orbit of Rosetta, and as such very sensitive to the Rosetta-comet distance $d_{CG}$, which was above 200~km for several months around perihelion. We have applied the method of \citet{vigren_predictions_2013} to estimate a maximum EUV absorption of only $0.8 (\pm0.1)$~percent by $\text{H}_2\text{O}$ molecules near perihelion at $d_{CG} = 330$~km (the attenuation was calculated along the Sun-Rosetta line assuming a spherically symmetric coma decaying in number density as $d_{CG}^{-2}$). In the same scenario but close to the comet surface the maximum EUV absorption is instead $ 70(\pm 7)$~percent at certain wavelengths, and as such not in disagreement with Alice results. Unfortunately there are no direct measurements of incident solar UV from Alice or other Rosetta instruments, and we lack an absolute measure of the incident solar UV at Rosetta to compare with the RPCLAP photoemission.

\subsection{Attenuation by cometary dust} \label{sec:disc_dust}

Another possible source of EUV extinction would be scattering and absorption of cometary dust grains. Studies on interstellar dust by \citet{cruise_extreme-ultraviolet_1993} and \citet{kolokolova_physical_2004} indicate that EUV scattering by micrograins and nanograins can be significant. {Even so, the large dust grains mainly observed by the GIADA, COSIMA and MIDAS dust instruments \citep{rotundi_dust_2015,fulle_density_2015,bentley_aggregate_2016,hilchenbach_comet_2016} could not, for the amounts reported, provide sufficient surface area for our inferred UV decrease of 50~percent at perihelion. %{\color{red} Delete and switch with green paragraph? However, the particles investigated with the MIDAS instrument all show agglomerate character with subunit sizes down to the nanometer scale. Therefore, if the dust particles would undergo a process like fragmentation or erosion that leads to the creation of smaller grains, their total scattering area would increase and more significant attenuation in UV may result. If the nanograins would only be created farther away from the nucleus, this would explain why we do not see a large UV extinction difference for different Rosetta-comet distances e.g. between 200 and 1000~km in Oct 2015, and would also be in line with the detections of nanograins at comet Halley}~\citep{utterback_attogram_1990}. %and could explain the herein presented data.
However, the particles investigated with the MIDAS instrument all show agglomerate character with subunit sizes down to the nanometer scale. %Anders wanted this comment here, to reference burch earlier. But I feel its a bit out of place:
%In addition,~\citet{burch_observation_2015} presented an event where data from the Ion and Electron Sensor (IES) on Rosetta showed signatures expected of nanograins, so such may at least at times be present also at Rosetta position.
%In addition, we note that there are direct detections of nanograins \citep{dellacorte_shining_2015,burch_observation_2015,gombosi_negatively_2015}, where ~\citet{dellacorte_shining_2015} note higher flux of submicron grains from the subsolar point than from the nucleus.
%Nanograins were also occasionally detected by the Ion and Electron Spectrometer IES~\citep{burch_observation_2015} and seen to flow mainly in the antisunward direction. In addition,  GIADA noted a three times higher flux of submicron dust particles in the antisolar direction than the flux coming directly from the nucleus~\citep{dellacorte_shining_2015}. \citet{gombosi_negatively_2015}  modelled the influence of radiation pressure on grain motion, suggesting nanograins originate from larger grains emitted by the comet and fragmenting at distances of several thousand km sunward of the nucleus.
%Similarly, the dust collection of the fast Stardust fly-by at comet 81P/Wild demonstrate the existence of agglomerates with components as small as tens of nanometres~\citep{horz_impact_2006}. 
It is therefore conceivable that cometary dust particles could undergo a process like fragmentation or erosion that leads to the release of their constituent (nano)grains, thus increasing the total surface area and more significant attenuation in UV may result. %The increased scattering area of the smaller grains would then lead to a more significant attenuation in UV. 
%If the nanograins would be created farther away from the nucleus, this would explain why we do not see a large UV extinction difference for different Rosetta-comet distances e.g. between 200 and 1000~km in Oct 2015, and would also be in line with the detections of nanograins at comet Halley}~\citep{utterback_attogram_1990}.

If the nanograin production would be most efficient farther away than Rosetta's position from the comet nucleus, this would explain the absence of a large UV extinction difference for different Rosetta-comet distances (that were, e.g., between 200 and 1000~km in Oct 2015). It is notable that the first direct detections of nanograins at a comet were made during the fly-by at 1P/Halley, which covered distances that were mainly larger than the general Rosetta-comet distance, and that there are measurements suggesting high densities of nanograins at immense (in the order of 10$^6$ km) comet distances~\citep{utterback_attogram_1990}. 

From Rosetta at 67P, nanograins were occasionally detected by the Ion and Electron Spectrometer IES~\citep{burch_observation_2015} and seen to flow mainly in the antisunward direction. In addition, the GIADA dust detector noted a three times higher flux of submicron dust particles in the antisolar direction than the flux coming directly from the nucleus ~\citep{dellacorte_shining_2015}. \citet{gombosi_negatively_2015} modelled the influence of radiation pressure on grain motion, suggesting nanograins originate from larger grains emitted by the comet and fragmenting at distances of several (tens of?) thousand km sunward of the nucleus. The radiation pressure drives them back toward the comet as seen by IES and GIADA, also yielding a significant nanograin column density in the sunward direction from Rosetta as suggested in our scenario.

%Thus, the existence of nanograins at far comet distances could be in good agreement with our current understanding of cometary dust.

In the following we will test if our hypothesis of nanograins absorbing the EUV coming from the Sun can hold. We will check if fragmentation or erosion of a minor amount of dust particles at large distance from the nucleus can lead to a sufficient population of nanograins and estimate the necessary size to account for our observations. We will revisit Rosetta results to ensure their compatibility, and finally discuss the implications of a hypothetical existence of a nanograin population in a certain distance of the comet.

%Although the cross sections are not as well known as for H$_2$O photoabsorption, it could potentially be less sensitive for increasing $d_{CG}$, by fragmentation of the dust grains, increasing the available surface area of dust particles.

%At comet Halley \citet{utterback_attogram_1990} reports estimates of attogram grains accounting for several percent of the total mass loss, as well as indications of significant dust grain fragmentation. At 67P \citet{burch_observation_2015} have reported one observation of nanograins, arriving from the solar direction and thus likely to have been created by fragmentation at large distance. Also, as evident from studies on interstellar dust by \citet{cruise_extreme-ultraviolet_1993}and \citet{kolokolova_physical_2004}, EUV scattering by micrograins and nanograins can be significant, and may partially or completely explain our results.
% * <cyril.simon.wedlund@gmail.com> 2017-03-30T23:02:00.892Z:
% 
% > Although the cross sections are not as well known
% as for H$_2$O photoabsorption
% 
% ^ <frejon@irfu.se> 2017-03-31T12:44:44.679Z.

Consider large ($1-1000~\mu \text{m}$) grains being produced at the comet surface, ejected isotropically at some velocity $u$, and fragmenting or eroding into smaller particles of radius $a$ outside some distance $d_0$, scattering 100\% of their spherical geometric cross section such that the fraction of scattered light $\alpha_s$ is

\begin{equation}
\label{dust1}
\alpha_s \in \left[0, 1\right] = N\ \pi a^2 , 
%\alpha_s = N\ \pi a^2, \text{with } \alpha_s \in \left[0,\,1\right] ,
\end{equation}
% \begin{equation}
% \label{dust1}
% \alpha_s \in \left[0, 1\right] =
% 	\begin{cases}
% 	N \pi a^2 , & \text{for } \alpha<<1 \\
%     1- e^{-N \pi a^2} & \text{else}
% 	\end{cases}
% \end{equation}
where $N$ is the column density of spherical dust grains small enough for significant UV absorption. This assumes the fraction of scattered light, or the optical depth, to be small and for a more general case we note that Equation~\ref{dust1} becomes: 
\begin{equation}
\label{dust11}
-\ln(1-\alpha_s) =  N\, \pi a^2.
\end{equation}
If we let a fraction $f_{\text{frag}} < 1$ of the total mass of the dust cloud undergo fragmentation or erosion, then the total mass $M$ per area $A$ of dust in a column between Rosetta and the Sun becomes :

\[
\frac{M}{A} =  \frac{N\, m_g\,}{f_{\text{frag}}} =  \frac{4 \pi a^3\, \rho N}{3\,f_{\text{frag}} },
\]
where $m_g$ is mass of a dust grain fragmentation product of density $\rho_{\text{frag}}$. Inserting Equation~\ref{dust1} gives
\begin{equation}
\label{dust2}
\frac{M}{A} = \frac{4 a\, \rho\, \alpha_s}{3\,f_{\text{frag}} }.
\end{equation}
In a column from $d_0$ to the Sun, where we assume the dust grains have fragmented into small enough particles for significant UV scattering to take place, the mass per area is then

\begin{equation}
\label{dust3}
\frac{M}{A} = \int_{d_0}^{\text{Sun}} \rho_\text{vol}(r) dr  = \rho_0 \int_{d_0}^{\text{Sun}} \left(\frac{R}{r}\right)^2 dr  \approx  \frac{\rho_0\, R^2}{d_0}~,
\end{equation}
assuming an isotropic density distribution decreasing with $r^{-2}$ from the comet surface at $r = R$, where the volume average mass density of dust $\rho_\text{vol}(R)=\rho_0$.

Assuming a constant dust-to-gas mass ratio $C$, we can use the production rate $Q$ of water gas from \citet{hansen_evolution_2016}, to estimate $\rho_0$:

%Following reports of the dust to gas mass ratio ($C$) by\citet{snodgrass_perihelion_2016} and others, we can use the production rate $Q$ of water gas from \citet{hansen_empirical_2016,hansen_combining_2015}, to estimate $\rho_0$:

\begin{equation}
\label{dust4}
\rho_0 = \frac{m_g\, Q_\text{dust}}{4\pi\, R^2\, u} = \frac{ C\ m_\text{H$_2$O}\, Q_\text{H$_2$O}}{4\pi\, R^2\, u}
\end{equation}

%\frac{M}{A} = \frac{4}{3}\, a\, \alpha_s \rho. 

Finally, by combining Equations~\ref{dust2},~\ref{dust3} and~\ref{dust4} we obtain:
%\[
%\frac{4}{3}\ a\rho\, \alpha_s  = \frac{C\, %m_\text{H$_2$O}\,Q_\text{H$_2$O}\, R^2}{4\pi R^2 u d_0},
%\]
\[
\frac{4 a\, \rho\, \alpha_s}{3  f_{\text{frag}} }
 = \frac{C\, m_\text{H$_2$O}\,Q_\text{H$_2$O}\, R^2}{4\pi R^2 u d_0},
\]
solving for $a$ yields
\begin{equation}
\label{dust5}
a = \frac{ 3\ f_{\text{frag}}\, C\, m_\text{H$_2$O}\, Q_\text{H$_2$O}}{16\pi\, u\, d_0\,  \alpha_s\, \rho  }.
\end{equation}

%At perihelion, for a dust grain with outflow velocity and density in the middle of the range reported by \citet{fulle_density_2015} ($u = 5$~m/s, $\rho = 1500$~kgm$^{-3}$), dust to gas mass ratio $C=5$ \citep{snodgrass_perihelion_2016}, $Q_{H_2O}=10^{28}$~s$^{-1}$, absorbing $\alpha_s = 0.5$ of incoming EUV in a column from $d_0$ = 1000~km to the Sun, we estimate the dust grain radius to be $\approx$ 17~nm.

At perihelion for a dust grain with average outflow velocity $u = 3$~m/s as reported by \citet{fulle_density_2015}, dust bulk density $\rho = 800$~kgm$^{-3}$ \citep{rotundi_dust_2015}, dust-to-gas mass ratio $C=5$ \citep{snodgrass_perihelion_2016}, $Q_\text{H$_2$O}= 3.5 \times 10^{28}$~s$^{-1}$ \citep{hansen_evolution_2016}, and letting   $f_{\text{frag}}=10~\%$ of the dust mass fragment and absorb $\alpha_s = 50~\%$ of incoming EUV in a column from $d_0$ = 1000~km to the Sun, we estimate the dust grain radius to be $\approx$ 19~nm.

%$107~nm <-old

A spherical dust grain with radius of 19~nm would according to \citet{skolnik_introduction_1981} scatter 100 percent of its geometric cross-section of light in wavelength of $2 \pi a = $~119~nm and below, and is as such on the correct length scale for attenuation in RPC-LAP wavelengths. The above considerations are of course very rough, resting on a series of assumptions. Nevertheless they show that the above hypothesis cannot be ruled out directly. An obvious simplification in the model is the singular size of the disintegration product. A size distribution, although useful, should be cemented in a firm understanding of the disintegration process involved, including disintegration products and forces, which we do not pretend to have. However, if the simple model works for a singular size of grains, then it will also work for some distribution of grains. Furthermore, due to our lack of physical model for the disintegration, $d_0$ is more or less a free parameter. However, we chose a value consistent with (1) a negligible decrease of attenuation even during the day-side excursion (Sep-Oct 2015, up to 1000~km sunward), (2) remote observations~\citep{boenhardt_wendelstein_2016} of 67P dust, discussed in more detail below, and (3) still much less than the apex distance for dust grains as investigated by \citet{gombosi_negatively_2015}.

%If our model works for a singular size of grains, it should also work for some size distribution, which should be cemented in a firm understanding of the disintegration process involved, which we do not pretend to have. 

%This line below is removed. Why focus on a single estimate, when all estimates from the paper would have some uncertainty.
%We note also that in the velocities reported from \citet{fulle_density_2015}, significant dust deceleration from the negative spacecraft potential is speculated, resulting in our velocity estimate possibly being too low. 

To compare these results with other Rosetta observations it should be noted that the presented RPC-LAP measurements were taken by remotely sampling the Rosetta-Sun environment. Rosetta carried a variety of other remote instruments, although most investigated the vicinity of the comet nucleus and rarely sampled sunward. Additionally, the different spectral ranges of the instruments further impede a detection of UV-extinction as, e.g., the scientific camera system on-board OSIRIS is sensitive in the range of 250 - 1000 nm \citep{keller_osiris_2007}), for which the geometric scattering efficiency of 19~nm sized nanograins would decrease to 4 - 0.1\% \citep{skolnik_introduction_1981} and thus possibly escape detection.
%{\color{green} added this section above here, since I don't want the $u$ and $d_0$ discussion to be too far away from the dust grain radius.}

As stated in the beginning of this section, the amounts reported in direct observations of large dust grains cannot directly account for the inferred UV attenuation. Our model thus relies on a mechanism of fragmentation or erosion of large grains at larger distances than the typical Rosetta-comet distance. One such mechanism would be erosion and/or evaporation of gluing material~\citep{boenhardt_wendelstein_2016, lasue_interplanetary_2007}. As this process would be most effective for periods with high solar radiation it would readily account for the strong UV absorption during perihelion whilst fading to absence for increasing comet-Sun distances. This scenario is also in agreement with remote observations of comet 67P from Earth by~\citet{boenhardt_wendelstein_2016}, where their observational data suggests dust fragmentation at large comet distances, in particular for perihelion when the vicinity of the comet to the Sun facilitates dust heating and thus material degradation. Furthermore, the modelling work of~\citet{gombosi_negatively_2015} shows that dust particles ejected sunward may be deflected by solar radiation at a comet distance of some thousand kilometres, and suggest particle fragmentation close to their turn-around point. Finally, as comets are speculated to be a source of nanodust in our Solar System~\citep{mann_comets_2017}, the herein presented hypothesis might aid the understanding of the comet contribution  to the Solar System dust.

\section{Conclusions}
\label{sec:conc}
We have presented estimates of the Langmuir Probe photoemission current using three different methods of which one is, to our knowledge, new. 
All three methods agree very well on global and small scales and enables the use of the Langmuir Probe as an ultraviolet photo-diode on Rosetta. The three methods are further validated by the use of theoretical estimates of the Langmuir probe photoelectron emission using EUV measurements from two other spacecraft, and a suitable estimate for the photoelectron yield of the titanium probe, which agrees very well on the start and end of the mission as well as small scale fluctuation in solar sidereal rotation frequencies. The results in this paper can be used to estimate the solar EUV intensity at the Rosetta position, as well as cataloguing flares. We also report a significant current discrepancy from our measured values to the EUV estimates around perihelion, correlating with high cometary activity. Although there are many sources of errors of any method individually, only contamination is common between all three methods, of which no evidence has been found. Attenuation by gas emitted from the nucleus cannot explain the decreased photoemission. However, a test model of attenuation by erosion or fragmentation of dust creating grains of tens of nanometers far from the comet is found to be consistent with observations.

%Other candidate theories are discussed to explain the discrepancy with attenuation due to dust and gas is considered the most likely. A test model of attenuation by fragmented dust grains of tens of nanometers far from the comet is found to be consistent with observations on other instruments.

% * <cyril.simon.wedlund@gmail.com> 2017-03-30T23:06:42.846Z:
% 
% > titanium
% titanium can be in lower case I think. Check everywhere in the text. Talking about that I homogenised MAVEN --> MAVEN everywhere in the text. Still in certain legends of the figures, it is written MAVEN. Maybe to change if you have time.
% 
% ^.

%The results in this paper can be used to estimate the solar EUV intensity at the Rosetta position, as well as cataloguing flares. We suggest future work in confirming and estimating ultraviolet attenuation by neutral gas species such as $\text{H}_2\text{O}$ and $\text{CO}_2$, as well as cometary dust.

% * <cyril.simon.wedlund@gmail.com> 2017-03-30T23:16:16.709Z:
% 
% > as well as cataloguing flares
% This is an excellent idea, Lei has been also looking into that for a few months, let's keep each other updated.
% 
% ^.

%The comet is made of bubbles.

\section*{Acknowledgements}
Rosetta is a European Space Agency (ESA) mission with contributions from its member states and the National Aeronautics and Space Administration (NASA). This work has made use of the AMDA and RPC Quicklook database, provided by a collaboration between the Centre de Donn{\'e}es de la Physique des Plasmas (CDPP) (supported by CNRS, CNES, Observatoire de Paris and Universit{\'e} Paul Sabatier, Toulouse), and Imperial College London (supported by the UK Science and Technology Facilities Council).
Support by the Swedish National Space board is acknowledged, including SNSB contracts 109/12, 135/13, 166/14 and 168/15. The work at the University of Oslo is partly supported by the Norwegian Research Council, grant number 240000. The contribution from the Space Research Institute of the Austrian Academie of Sciences was made possible by the funding of the Austrian Science Funds FWF P 28100-N36.
%%%%%%%%%%%%%%%%%%%%%%%%%%%%%%%%%%%%%%%%%%%%%%%%%%

%%%%%%%%%%%%%%%%%%%% REFERENCES %%%%%%%%%%%%%%%%%%

% The best way to enter references is to use BibTeX:

\bibliographystyle{mnras}
%\bibliography{Zotero2017} 
\bibliography{Zotero_bibtex_v4} 

%!!!!!!!
%If you need a reference, copy paste the bibtext below and I'll add it to the Zotero bib fil. //F
%!!!!!!!!!!

%%%%%%%%%%%%%%%%%%%%%%%%%%%%%%%%%%%%%%%%%%%%%%%%%%

%%%%%%%%%%%%%%%%% APPENDICES %%%%%%%%%%%%%%%%%%%%%

\appendix

\section{Probe photoemission from analysis of multiple sweeps, supplementary material }

We investigate the accuracy of the multiple sweep analysis method with synthetic data for a range of plasma parameters. To model the total current, we use Equations~\ref{eq:electron},~\ref{eq:ion}, and~\ref{eq:photoyield}. While in reality the noise level of the instrument is $\pm 0.5$~nA, we do not add noise to our modelled current since we focus on the ideal limits of this method. The current response is modelled for $V_S = -10$~V, $V_b \in \left[-30,30\right]$~V, fixed electron temperature at $T_e = 5$~eV, and varying ion velocity $u$ and plasma density $n$. The slope of $I_{tot}$ is then found by fitting a linear function to the synthetic current for $V_p \in \left[-40,-30\right]V$ and $I_{tot}$ is taken in the middle of that range from the fitted function, i.e $I_{tot}\left(V_p = -35 ~\text{V}\right)$. In Figure~\ref{fig:error_theory} we show results from the synthetic data for three different ion velocities, $u$ = 0.1~km/s, 50~km/s, and 350~km/s. The solid lines are calculated with Equation~\ref{eq:slope_v_current} for $I_{tot} \in [-50,0] ~nA$ for given ion velocity $u$. The important difference between the data points found from synthetic data and Equation~\ref{eq:slope_v_current} is that the synthetic data includes an electron current. The increasing values of $dI_{tot}/dV_p$ and $I_{tot}$ for the data points corresponds to increasing plasma density. One can see from the difference between the solid lines and the data points in Figure~\ref{fig:error_theory} that the electron contribution becomes increasingly important for higher plasma density. Consequently, the slope can be overestimated due to the electron current. In the case of Figure~\ref{fig:error_theory}, extrapolating data with a single $k$ would yield a large error in estimating $I_{ph0}$. In reality, the ion velocities would not change from 0.1~km/s to 350~km/s within a few sweeps.

To simulate more realistic plasma variations around comet 67P from sweep to sweep, we model the total current to the probe for a range of different parameters observed by RPCLAP. As can be seen from Equations~\ref{eq:slope_v_current}, and~\ref{eq:ion}, the relevant plasma parameters are the plasma density and the ion energy. Since $I_e$ is exponentially decreasing for $V_p < 0$ it is expected that some contributions from $I_e$ will affect our results, depending on the range of $V_p$ we fit a linear function to. If the fitting range is not sufficiently below 0 the electron current might not be negligible. Hence, to investigate the error source attributed to electron currents, we change the range of $V_p$ to which the linear function is fitted to against the synthetic $I_{tot}$. Furthermore, from \ref{eq:slope_v_current}, it is expected that with smaller ion velocity variations $\Delta u$ between the sweeps this method is more accurate, and thus, we will also consider different $\Delta u$ in the analysis. 

We take the plasma density in the range of $n \in \left[0,~4000\right]$~cm$^{-3}$, the ion velocity in the range of $u \in \left[1,~5\right]$~km/s, and the electron temperature to be constant at $T_e = $~5~eV. The range of $n$ is subdivided into 35 equispaced intervals, while $u$ is subdivided into two different equispaced intervals - with 35, and 105 intervals. Hence, $\Delta n \simeq 117$~cm$^{-3}$ is constant, and $\Delta u \simeq 117 ~\text{and} ~38$~m/s respectively between the sweeps. Thus, creating two different grids with the size of  $35 \times 35$ and $35 \times 105$. Within each grid square 9 different $I_{tot}$ are created - one for each combination of $n$ and $u$. For each $I_{tot}$, two different linear fits are fitted to the data - one fit for each $V_p$-range. 

The $I_{ph0}$ dataset from the multiple sweep method is made possible due to a a dynamic fitting routine of $V_S$ ( as previously mentioned in section~\ref{sec:method1}) such that the $V_p$ range to estimate the slope and offset is also dynamic for each sweep, with most of the dataset having $V_p$ ranges between 10 and 15~V, and is also the chosen $\Delta V_p $-ranges in the synthetic sweeps. The slope $dI_{tot}/dV_{P}$ is found from the fitted linear function, and $I_{tot}$ is found in the middle of the $V_p$-range from the same fitted function. $dI_{tot}/dV_{P}$ versus $I_{tot}$ is then plotted, and a linear function is fitted these data-points to extract $I_{ph0}$, as illustrated in Figure~\ref{fig:error_theory}. 

In Figure~\ref{fig:error_plot} we plot the error of $I_{ph0}$ found by this method as compared to the $I_{ph0} = -10$~nA we used to create the synthetic data. Panels a) and c) has a range of $V_p \in \left[-40, -30\right]$~V, and panels b) and d) a range of $V_p \in \left[-40, -25\right]$~V. We observe that this method remains accurate for lower plasma densities, even for a larger fitting range. The errors increase for increasing density, especially for fitting ranges closer to $V_S$, due to the exponential part of the electron current. The error decreases for smaller ion velocity variations between the sweeps, i.e, with smaller $\Delta u$, as is illustrated by the difference between panel b and d of Figure~\ref{fig:error_plot}. 

\begin{figure}
	% To include a figure from a file named example.*
	% Allowable file formats are eps or ps if compiling using latex
	% or pdf, png, jpg if compiling using pdflatex
	\includegraphics[width=1.0\columnwidth]{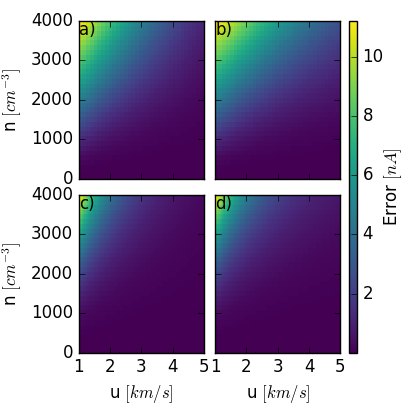}
    \caption{The error in $I_{ph0}$ for two different levels of velocity variation $\Delta u$ and two different ranges of $V_p$. $V_p$ is varied for each column so that panels a and c has $V_p \in \left[-40, -30\right]$~V. Panels b and d has $V_p \in \left[-40, -25\right]$~V. Panels a and b has a velocity variation of $\Delta u \simeq 117$~m/s between the sweeps used for each grid square and for panel c and d $\Delta u \simeq  38$~m/s.}    
    \label{fig:error_plot}
\end{figure}

It is evident from Figure~\ref{fig:error_plot} that this method is robust for most of the plasma conditions around 67P. A smaller range of $V_p$ would reduce noise in the method, although instrumental noise provides a lower bound of the range of $V_p$ for a good fit to be found. Further errors might be introduced due to heavy fluctuations of the ion energy in a very dense plasma. As previously mentioned, we do not expect this method to be sensitive to $I_{SEEP}$. If anything it would mitigate the electron retardation current influence on our $I_{ph0}$ estimate, which we consider is our largest error source, due to the opposite sign and slope of the secondary electron current.

\label{sec:app_oslo}

%\section{Contamination}
%\label{sec:app_cont}

%%%%%%%%%%%%%%%%%%%%%%%%%%%%%%%%%%%%%%%%%%%%%%%%%%

% Don't change these lines
\bsp	% typesetting comment
\label{lastpage}
\end{document}